\documentclass[twocolumn]{aastex63}

\usepackage{booktabs}

\submitjournal{ApJL}

\shorttitle{Cold giant planets evaporated by hot white dwarfs} 
\shortauthors{Schreiber et al.}

\begin{document}

\author{Matthias R. Schreiber}
\affil{Instituto de F{\'i}sica y Astronom{\'i}a, Universidad de Valpara{\'i}so, Valpara{\'i}so, Chile; matthias.schreiber@uv.cl}
\affil{Millennium Nucleus for Planet Formation, NPF, Universidad de Valpara{\'i}so, Valpara{\'i}so, Chile}

\author{Boris T. G\"ansicke}
\affil{Department of Physics, University of Warwick, Coventry CV4 7AL, UK}
\affil{Centre for Exoplanets and Habitability, University of Warwick, Coventry CV4 7AL, UK}

\author{Odette Toloza}
\affil{Department of Physics, University of Warwick, Coventry CV4 7AL, UK}

\author{Mercedes-S. Hernandez} 
\affil{Instituto de F{\'i}sica y Astronom{\'i}a, Universidad de Valpara{\'i}so, Valpara{\'i}so, Chile; matthias.schreiber@uv.cl}

\author{Felipe Lagos}
\affil{Instituto de F{\'i}sica y Astronom{\'i}a, Universidad de Valpara{\'i}so, Valpara{\'i}so, Chile; matthias.schreiber@uv.cl}
\affil{Millennium Nucleus for Planet Formation, NPF, Universidad de Valpara{\'i}so, Valpara{\'i}so, Chile}

\title{Cold giant planets evaporated by hot white dwarfs}


\begin{abstract}
Atmospheric escape from close-in Neptunes and hot Jupiters around sun-like stars driven by extreme ultraviolet (EUV) irradiation plays an important role in the evolution of exo-planets and in shaping their ensemble properties. Intermediate and low mass stars are brightest at EUV wavelengths at the very end of their lives, after they have expelled their envelopes and evolved into hot white dwarfs. Yet the effect of the intense EUV irradiation of giant planets orbiting young white dwarfs has not been assessed. We show that the giant planets in the solar system will experience significant hydrodynamic escape caused by the EUV irradiation from the white dwarf left behind by the Sun. A fraction of the evaporated volatiles will be accreted by the solar white dwarf, resulting in detectable photospheric absorption lines. As a large number of the currently known extra-solar giant planets  will survive the metamorphosis of their host stars into white dwarfs, observational signatures of accretion from evaporating planetary atmospheres are expected to be common. In fact, one third of the known hot single white dwarfs show photospheric absorption lines of volatile elements, which we argue are indicative of ongoing accretion from evaporating planets. The fraction of volatile contaminated hot white dwarfs strongly decreases as they cool. We show that accretion from evaporating planetary atmospheres naturally explains this temperature dependence if more than 50 per cent of hot white dwarfs still host giant planets. 
\end{abstract}

\section{Introduction}
\noindent
Atmospheric escape has been a key topic in exo-planet research since \cite{mayor+queloz95-1} speculated that the hot Jupiter 51\,Peg\,b might be a radiatively stripped brown dwarf. Direct evidence for the escape of atmospheric hydrogen was 
first 
established by \textit{Hubble Space Telescope} Ly$\alpha$ transit spectroscopy of the hot Jupiter HD\,209458 \citep{vidal-madjaretal03-1}, later for close-in Neptunes \citep{ehrenreichetal15-1, bourrieretal18-1}, and possibly  super-Earths \citep{ehrenreichetal12-1}. Outflowing atmospheres have also been detected from large transit depths in the far-ultraviolet resonance lines of  \ion{O}{1} and \ion{C}{2} \citep{vidal-madjaretal04-1,ben-jaffel+ballester13-1}. 

This atmospheric escape 
%from planets with hydrogen-rich atmospheres on close orbits around main sequence stars 
is driven by high-energy irradiation. Extreme ultraviolet (EUV, $\lambda=100-912$\AA) photons have sufficient energy to directly 
ionize hydrogen atoms ($h\nu=13.6$\,eV). A significant 
fraction of the energy of the generated photoelectrons is converted into heat as radiative cooling is inefficient. This drives the atmosphere out of hydrostatic equilibrium into a transonic flow \citep{parker64-1,tianetal05-1}. If the resulting escape flow of hydrogen is large enough, drag forces %will 
carry along heavier constituents of the atmosphere, as indicated by the observations \citep{yelleetal08-1, ben-jaffel+ballester13-1, vidal-madjaretal04-1}. 

As the mass loss rate increases with the amount of incident ionizing EUV radiation, hydrodynamic escape is expected to be important especially for planets around stars younger than $\simeq100$ million years, which have EUV fluxes 100--1000 times that of the present-day Sun \citep{tuetal15-1,ayres97-1}. Planet evaporation during these early phases of large host-star EUV luminosities is thought to explain both the dearth of Neptune mass planets with orbital periods shorter than $\sim$\,10 days, referred to as the ``warm Neptune desert'', as well as 
the low occurrence of planets with radii of $\simeq\,2$R$_\mathrm{Earth}$ at separations of 0.03--0.1\,au, usually termed the ``the evaporation valley'' \citep{owen+wu17-1}. 

One phase of stellar evolution where the large EUV luminosity of the host stars can drive significant atmospheric escape from giant planets has been overlooked so far: the very 
last stages of the evolution of planetary systems when the host star will have evolved into a hot white dwarf. The survival of planetary systems into the white dwarf stage is demonstrated by the detection of the debris from tidally 
disrupted planetesimals \citep{jura03-1, zuckermanetal03-1, koesteretal14-1}. Here we show that the EUV emission from hot white dwarfs may have far-reaching consequences for the future of the solar system as well as important implications for our understanding of metal pollution in hot white dwarfs. 

\begin{figure}
\includegraphics[angle=90,width=\columnwidth]{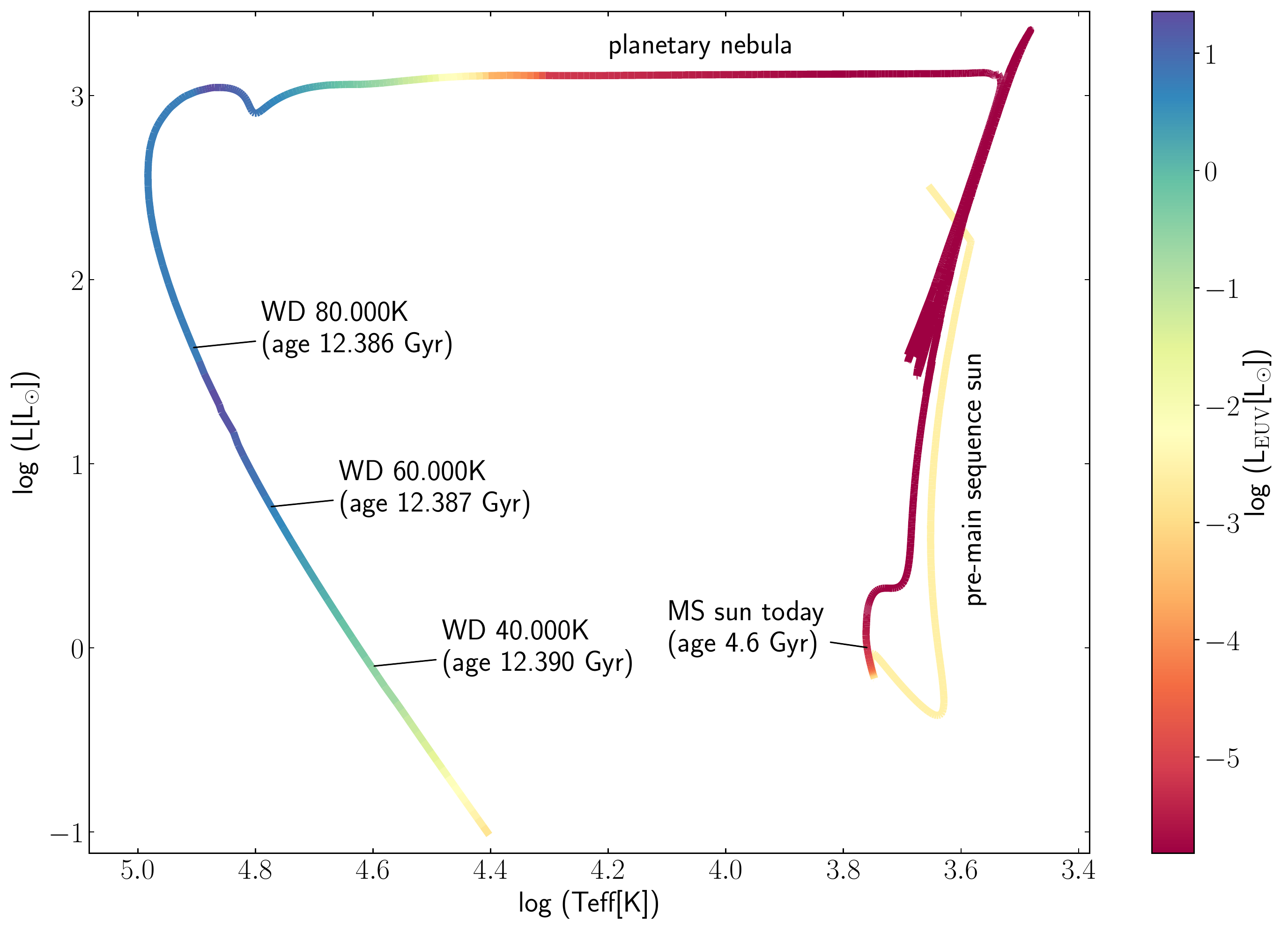}
\caption{\label{f-hrd}  The evolution of the EUV luminosity of the Sun across the Hertzsprung-Russel diagram. During the pre-main sequence phase, the EUV luminosity reaches $\simeq0.1-0.5$~per cent of the solar luminosity, and drops significantly during the main sequence and the giant branches. However, once the star expelled its envelope during the planetary nebular phase, its hot degenerate core~--~the white dwarf~--~is exposed, which for a few Myrs emits almost all its radiation in the EUV, resulting in $L_\mathrm{EUV}\gg L_\odot$. }
\end{figure}

\section{The future of the solar system}
\noindent
The Sun will slowly lose about half of its mass during the evolution through the giant branches, causing the giant planets to spiral out~--~while remaining gravitationally bound~--~to roughly twice their current orbital separations \citep{duncan+lissauer98-1}. The solar white dwarf will initially have an effective temperature of $\ga100\,000$\,K and cool quasi-exponentially to $\simeq55\,000$\,K  within 1\,Myr, and to $\simeq30\,000$\,K within 10\,Myr \citep{fontaineetal01-1}. Consequently, the solar white dwarf remains a luminous EUV source for several Myr.  
%\fix{sollten wir ``several ten Myrs'' sagen? Und ist die Einheit ``Myrs'' oder ``Myr''? 100\,cm und nicht 100\,cms?}

\subsection{The EUV luminosity of the solar white dwarf\label{s-solsys}}
\noindent
To establish the EUV luminosity of the solar white dwarf, we computed a grid of synthetic spectra extending from 10\,\AA\ to 25\,000\,\AA\ %\fix{... ein Pedant k\"onnte fragen warum wir hier 10\,\AA\ sagen, aber weiter oben und unten 100\,\AA} 
using the model atmosphere code of \citet{koester10-1}. The atmospheric input parameters are the effective temperature ($T_\mathrm{eff}$) and surface gravity ($\log g$) of the white dwarf. For a fixed mass, the white dwarf slightly contracts as it cools, and hence its surface gravity increases with time. We therefore interpolated the cooling tracks of \citet{holberg+bergeron06-1} for $M_\mathrm{wd}=0.5\,\mathrm{M}_\odot$ to determine $T_\mathrm{eff}$, $\log g$, the white dwarf radius and its cooling age. 

We computed the synthetic spectra assuming  pure hydrogen atmospheres. 
The presence of metals in the atmosphere will modify the emerging spectral energy distribution, in particular iron which has a large number of absorption lines in the EUV. This additional opacity is blocking the outgoing flux at short wavelengths, re-distributing the energy to longer wavelengths. However, even with trace-metals in their atmospheres, hot white dwarfs remain luminous EUV sources (see Fig.\,5 of \citealt{chayeretal95-2}).

The evolution of the EUV luminosity of the Sun from the pre-main sequence into the white dwarf cooling track\footnote{The stellar evolution track was calculated with \textsc{MESA}\citep{paxtonetal11-1}} 
%using the \textsc{1M\_pre\_ms\_to\_wd} inlist} 
is illustrated in Fig.\,\ref{f-hrd}. We assumed $L_\mathrm{EUV}=10^{31}\,\mathrm{erg\,s^{-1}}$ for the pre-main sequence phase, which most 
likely represents an upper limit \citep{tuetal15-1}. For the EUV luminosity on the main sequence we used Eq.(4) of \citet{sanz-forcadaetal11-1}
%which is based on observations of sun-like stars. 
and extended this relation to the giant-branch which should result in approximately correct EUV luminosities \citep{pizzolatoetal00-1}. For simplicity we interpolated the EUV flux from the planetary nebula phase to our first white dwarf model. The EUV flux for the white dwarf phase was determined integrating the synthetic spectra from 100\,\AA\ to 912\,\AA. For the initial few Myr, the EUV luminosity of the solar white dwarf will exceed that of the present-day Sun by $10^5-10^6$, and that of the young, chromospherically active Sun by $10^2-10^3$. The extreme EUV luminosity of the solar white dwarf will drive atmospheric mass loss from the giant planets that survived the metamorphosis of the Sun. 

\subsection{Photo-evaporation of the giant planets}

\noindent
Motivated by the detections of extended atmospheres of hot Jupiters and close-in Neptunes, 
%1D 
hydrodynamic calculations of EUV irradiated hydrogen atmospheres have been developed throughout the last decade \citep{yelleetal08-1,murray-clayetal09-1,owen+alvarez16-1}. According to these models, the hydrodynamic wind generated by EUV radiation can be separated in two different regimes. At lower fluxes, lower densities result in relatively long recombination times, adiabatic expansion dominates the gas cooling, and the wind mass loss rate is proportional to the incident EUV flux. This regime is called \textit{energy limited}. 
At high fluxes, the recombination time scale decreases and the wind mass-loss scales approximately with the square root of the incident EUV flux \citep[see][for details]{owen19-1}. This regime is called \textit{recombination limited}. 
Many variants of equations describing energy limited hydrodynamic escape appear in the literature, and we use the expression 
\begin{equation}
    \dot{M}_{\mathrm{elim}}= \frac{\beta \pi F_{\mathrm{EUV}} R_{\mathrm{P}}^3}{G M_{\mathrm{P}}}
\end{equation}
where $M_{\mathrm{P}}$ and $R_{\mathrm{P}}$ are the mass and radius of the planet, $F_{\mathrm{EUV}}$ is the incident EUV flux, $G$ is the gravitational constant, and $\beta$ is an efficiency parameter which is typically adopted to be $\simeq0.2$ \citep{murray-clayetal09-1}. 
%The planets that we consider in our analysis are located at orbital separations $>5$\,au from the white dwarf and the Roche-lobe correction term suggested by \cite{erkaevetal07-1} is therefore negligible. 

The transition from the recombination limited regime into the energy limited regime occurs over a range of limiting fluxes and depends on the mass and radius of the planet \citep{owen+alvarez16-1}. For simplicity, we assume a flux of $F_{\mathrm{EUV}}=10^4\,\mathrm{erg\,s^{-1}}$ for this transition, as suggested for the case of a hot Jupiter by \cite{murray-clayetal09-1}.
To ensure the robustness of our results, we repeated our analysis with several different prescriptions, e.g. a flux limit of $F_{\mathrm{EUV}}=10^3\,\mathrm{erg\,s^{-1}}$ for cold Jupiters, a much higher limit
%value 
of $F_{\mathrm{EUV}}=10^5\,\mathrm{erg\,s^{-1}}$ for Neptune mass planets 
%$($F_{\mathrm{EUV}}=10^5\,\mathrm{erg\,s^{-1}}$, \citealt{owen+alvarez16-1})
\citep{owen+alvarez16-1}, and a linear interpolation in between. Our results are not affected by the choice of the flux at which the transition occurs.

\begin{figure}
\includegraphics[angle=0,width=\columnwidth]{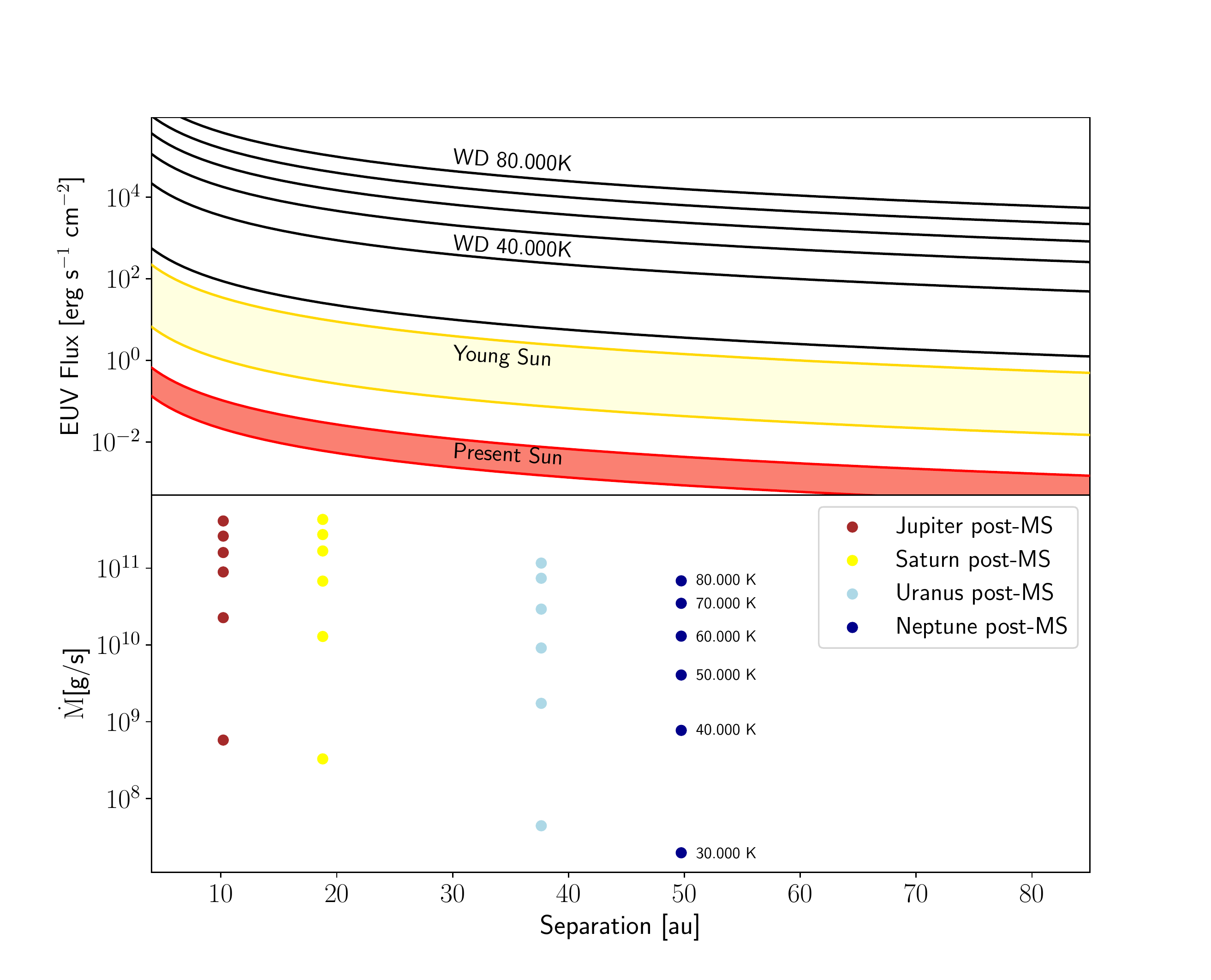}
\caption{\label{f-solsys} Top panel: comparison of the incident EUV flux per unit area onto giant planets as a function of the orbital separation. The irradiating flux from hot 
%and young 
white dwarfs ($80\,000-30\,000$\,K) can exceed those of the young Sun by many orders of magnitude. This 
%very 
intense EUV radiation can cause atmospheric escape in giant planets located at separations as large as $\simeq100$\,au. Bottom panel: estimated mass loss of the  giant planets in our solar system once the Sun has transformed into a white dwarf for temperatures of $80\,000-30\,000$\,K.  The mass loss rates of the giant planets orbiting the young solar white dwarf are comparable to those observed from exo-planets on close-in orbits around main-sequence stars.}
\end{figure}

Figure\,\ref{f-solsys} compares the incident EUV flux from the current Sun, the young and active Sun, and the solar white dwarf for temperatures in the range of 80\,000--30\,000\,K, as a function of the semi-major axis. The very large EUV irradiation from the white dwarf will drive mass loss rates of $\dot M\sim10^8-10^{11}\,\mathrm{g\,s^{-1}}$ from all giant planets in the solar system, despite their large orbital separations. These mass loss rates are comparable to those measured for hot Jupiters and warm Neptunes
%on close orbits around sun-like stars 
\citep[e.g.][]{vidal-madjaretal03-1,ehrenreichetal15-1}. While surprising at first, this result is a simple consequence of the large EUV luminosity of the white dwarf, $10^{5}-10^{6}$ that of the present-day Sun, which compensates for the $10^2-10^3$ times larger separation of the surviving planets compared to the short-period evaporating planets found around main-sequence stars. 

\subsection{Observational signatures of giant planet evaporation}

\noindent
Given their 
%extremely 
large surface gravity, white dwarfs rapidly undergo chemical stratification via gravitational settling \citep{schatzman48-1}, and their atmospheres are expected to be composed of the lightest elements, hydrogen or helium. Nevertheless, photospheric trace metals are commonly observed in the ultraviolet and optical spectra of cool ($T_\mathrm{eff}\lesssim20\,000$\,K) white
dwarfs \citep{zuckermanetal03-1, koesteretal14-1, jura+young14-1}, and it is now well-established that these white dwarfs accrete the debris of tidally disrupted planetesimals \textbf{\citep{jura03-1, veras16-1, farihi16-1}}. 

%It is unavoidable that 
The solar white dwarf will capture a fraction of the material escaping the atmospheres of the giant planets that survived the Sun's post-main sequence evolution. Taking into account that the flow is supersonic, the Hoyle-Littleton approach is applicable \citep{shapiro+lightman76-1,wang81-1}. 
The white dwarf will accrete material within a cylinder of radius $r_a=\psi 2GM/v_{\mathrm{rel}}$ with $\psi$ being a factor of order unity representing deviations from the Hoyle-Littleton prescription, and $v_{\mathrm{rel}}$ being the relative velocity between the white dwarf and the wind. The resulting accretion rate can then be written as \citep{shapiro+lightman76-1,wang81-1}: 
\begin{equation}
\dot{M}_{\mathrm{acc}}\simeq\pi\,r^2_a\rho(a) v_{\mathrm{rel}}(a)%\simeq \dot{M}_{\mathrm{evap}}/4.0
\end{equation}
where $a$ is the separation of the planet, and the relative velocity can be approximated as $v_{\mathrm{rel}}(a)=(v_{\mathrm{evap}}^2+v_{\mathrm{x}}^2)^{1/2}$
with $v_\mathrm{evap}$ being the wind velocity 
%ist die Geschwindigkeit die wir unten diskutieren...?} 
and $v_{\mathrm{x}}=2\,\pi\,a/P_{\mathrm{orb}}$. The density of the outflow 
%at the location of the white dwarf 
can be estimated assuming spherically symmetric mass loss of the planet, $\rho=\dot{M}_{\mathrm{evap}}/(4\pi a^2)$. For a given mass loss rate from the planet, the resulting mass accretion rate onto the white dwarf depends critically on $v_{\mathrm{rel}}$, thus on the wind velocity $v_\mathrm{evap}$, which we assumed to be constant. This quantity is largely unknown. 
Detailed models of hydrodynamic escape show that the escape velocity is typically reached at separations of a few times the radius of the planet \citep{tianetal05-1,
tripathietal15-1}. We here assume the wind velocity to be equal to the escape velocity at four times the planets radius and note that our main conclusions are not very sensitive to this assumption, i.e. varying the velocity by a factor of two does not lead to significant changes.

While the ionisation of hydrogen drives the hydrodynamic escape, only the heavier elements dragged with the flow will be accreted by the white dwarf: Ly$\alpha$ radiation pressure from hot white dwarfs substantially exceeds the gravitational attraction, and effectively prevents the inflow of hydrogen onto the white dwarf \citep{brownetal17-1}. As little is known regarding the exact composition of the evaporated material, we adopted solar metallicity i.e. a metal mass fraction of two per cent, for the calculation of the accretion rate of metals onto the white dwarf. 
%Consequently, the accretion rates of metals are assumed to be a factor of 50 lower than the total accretion rates given by Eq.(2). 

%For the heavier elements 
We find initial metal accretion rates onto the white dwarf of $\simeq10^8\,\mathrm{g\,s^{-1}}$
%immediately 
and a gradual decrease 
%after the white dwarf is formed and very hot ($\simeq100\,000$\,K), which gradually drop  
for 4\,Myr to $\simeq10^5\,\mathrm{g\,s^{-1}}$ as the white dwarf cools to $\simeq45\,000$\,K (Fig.\,3). For cooler white dwarfs, the peak of their spectral energy distribution shifts from the EUV into the ultraviolet, and hence both its efficiency at evaporating the giant planet atmospheres, and the resulting accretion rate onto the white dwarf drops more steeply. 

Given that white dwarf atmospheres are intrinsically devoid of metals, optical and ultraviolet spectroscopy is 
%extremely 
sensitive to the detection of traces of accreted planetary material, corresponding to accretion rates of as little as $10^5\,\mathrm{g\,s^{-1}}$ \citep{koesteretal14-1}. We conclude that the solar white dwarf will accrete significant amounts of volatiles from the evaporating giant planets and the accreted planetary material will be spectroscopically detectable by future generations of alien astronomers for several Myr. 
%million years. 

\begin{figure}
\includegraphics[angle=0,width=\columnwidth]{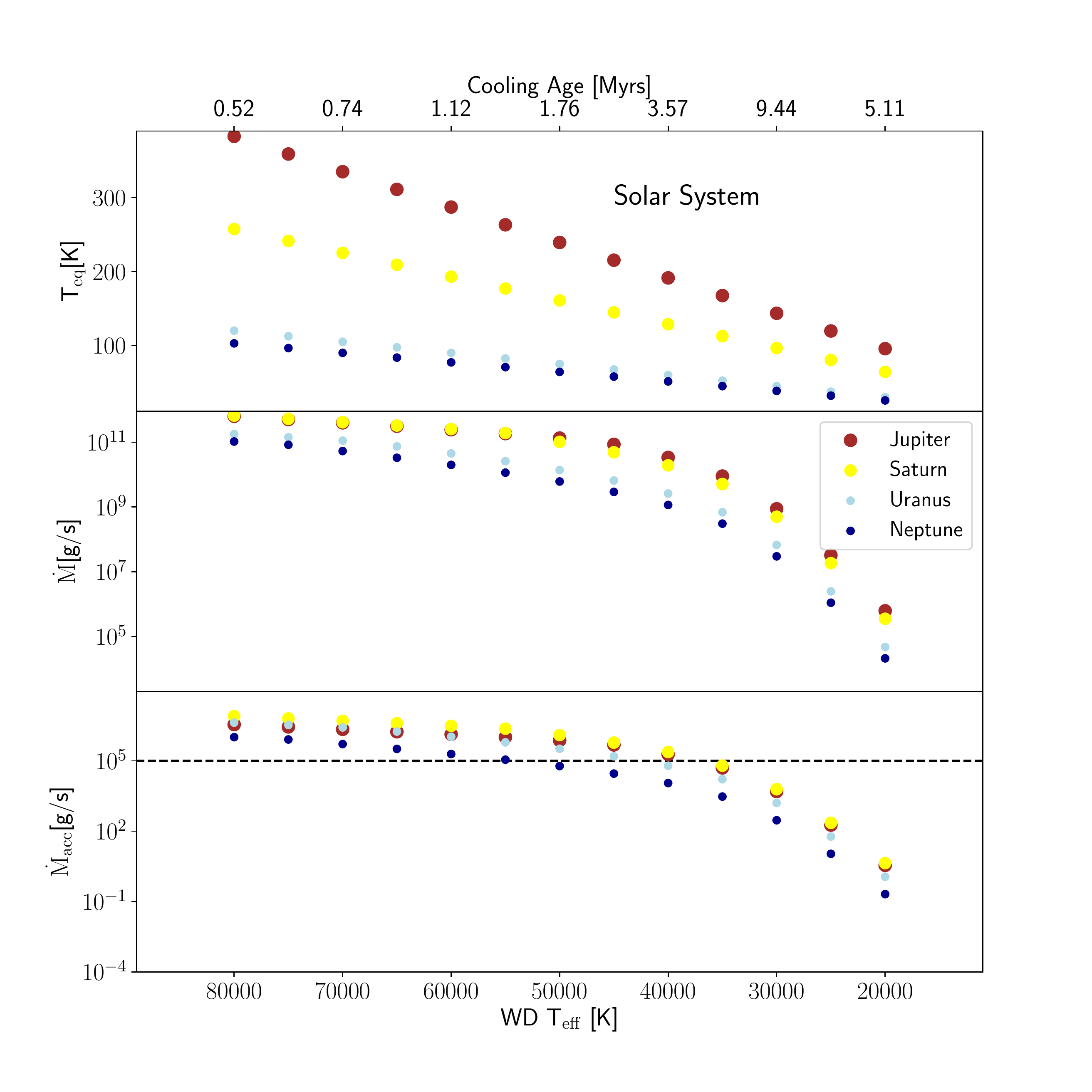}
\caption{\label{f-teff} Top panel: The equilibrium temperatures, mass loss rates, and accretion rates onto the solar white dwarf for the four giant planets as a function of its effective temperature. The corresponding cooling ages are given on the top axis. Given their large separations from their host white dwarf, the equilibrium temperatures of the planets are relatively low and monotonously decrease as the white dwarf cools. Middle panel: The mass loss rates driven by the EUV irradiation will remain high, $10^{11}-10^{9}\,\mathrm{g\,s^{-1}}$, for white dwarf temperatures exceeding $\simeq 45\,000$\,K and drop by $2-4$ orders of magnitude for cooler white dwarfs. Bottom Panel: The accretion rate of metals will detectable for $\dot{M}_{\mathrm{acc}}\ga 10^5\,\mathrm{g\,s^{-1}}$ (i.e. for $T_{\mathrm{eff}}\gtrsim 45\,000$\,K, dashed line) for $\simeq4$\,Myr after the formation of the white dwarf.}
\end{figure}

\begin{figure}
\includegraphics[angle=90,width=\columnwidth]{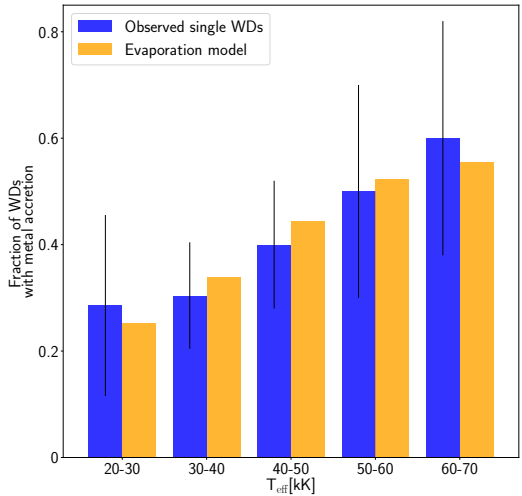}
\caption{\label{f-sim_obs_comp} Comparison between the observed fraction of photospheric trace metals detected in hot white dwarfs (blue, \citealt{barstowetal14-1})  with the predictions of our model population of evaporating giant planets (orange, assuming a planet occurrence rate of 65 per cent). In general the agreement is very good for planet occurrence rates exceeding $\simeq 50$~per cent.}
\end{figure}

\section{Photospheric volatiles at hot white dwarfs}

\noindent
The detection of photospheric metals in hot ($\gtrsim20\,000$\,K) white dwarfs is challenging from the ground, as few elements have sufficiently strong optical transitions. However, 
%early 
ultraviolet spectroscopy obtained with the \textit{International Ultraviolet Explorer} revealed strong photospheric metal lines in several hot white dwarfs \citep{bruhweiler+kondo81-1}. Subsequent studies with the \textit{Hubble Space Telescope} and the \textit{Far Ultraviolet Spectroscopic Explorer} (\textit{FUSE}) showed photospheric trace metals in many white dwarfs hotter than $\simeq50\,000$\,K \citep[e.g.][]{barstowetal03-1}.  

In contrast to cool white dwarfs, where the photospheric abundances reflect the rocky nature of the accreted material \citep{xuetal14-1}, these hot white dwarfs show large abundances of volatiles, in particular C, S and P. Whereas the interpretation of these observations is more complex than for cooler white dwarfs because of
%of the effect of
radiative forces counteracting gravitational settling, the consensus is that the abundances determined from the observations do not match the predictions of equilibrium radiative levitation theory \citep{chayeretal95-2}. In the most comprehensive far-ultraviolet study of hot white dwarfs, \citet{barstowetal14-1} detected photospheric C, P, or S in 33 out of 89 white dwarfs. \citet{barstowetal14-1} re-iterated that the observed pattern of abundances is incompatible with the predictions of radiative levitation, and concluded that accretion of planetary debris is the most likely explanation. The presence of reservoirs of this material is corroborated by the detection of circumstellar absorption lines \citep{dickinsonetal13-1}, which are only present in the spectra of hot white dwarfs that also show photospheric trace metals. 

%Here 
We argue that the volatiles detected among about a third of the hot ($T_\mathrm{eff}\ga20\,000$\,K) white dwarfs with adequate ultraviolet spectroscopy, and their observed increasing abundances with $T_\mathrm{eff}$, are the signature of evaporating giant planets (gas or ice giants planets from super-Earths to Jupiters).
%The implication of 
This hypothesis implies 
that a significant fraction of white dwarf progenitor stars hosted giant planets at separations large enough to survive their post-main sequence evolution. 

\section{Planet occurrences at large separations around A\&F stars}

\noindent
Most of the currently known white dwarfs had progenitors with masses of $1.5-2.5\,\mathrm{M}_{\odot}$ and spectral types F to A. 
%Whereas early radial velocity studies were inefficient at discovering planets around stars in this mass range, 
Direct imaging revealed the existence of Jupiter mass giant planets at larger separations ($5-100$\,au) around young A-type stars \citep[e.g.][]{maroisetal08-1,chauvinetal17-1}. Recent radial velocity studies indicate an increase in both the occurence rate and orbital separations of Jupiter-mass planets with increasing stellar mass \citep{borgnietetal19-1}. Surveys targeting sub-giant stars with $M\geq 1.6\,\mathrm{M}_{\odot}$, the so-called "retired A stars", confirm this trend: the giant planet occurrence increases with stellar mass up to $\simeq2\mathrm{M}_{\odot}$ \citep{ghezzietal18-1}. Just as in the solar system, these  giants planets will survive the evolution of their host stars into white dwarfs.

Currently the only method capable of detecting Neptune mass planets 
%with separations beyond the snow line, which 
that will survive the evolution of their host stars into white dwarfs is microlensing. Whereas microlensing surveys confirm the relatively low occurrence rates for Jupiter mass planets they consistently find large occurrence rates for Neptune and super-Earth mass planets. The fraction of bound planets at distances of $0.5-10$\,au from their host 
stars was found to 
be $17^{+6}_{-9}$ per cent in the mass range 0.3-10 M$_{\mathrm{Jup}}$ but significantly larger, $52^{+22}_{-29}$ and $62^{+35}_{-37}$ per cent, for Neptunes and super-Earths, respectively \citep{cassanetal12-1}. While this study relied on a small sample of events \citep{suzukietal16-1}, the main conclusions have been independently confirmed \citep[e.g.][]{shvartzvaldetal16-1}. 

\section{A planet population model}

\noindent
%\textbf{
Based on the observational evidence for a rising occurrence rate of Jupiter mass planets with increasing stellar mass, we extrapolated from the microlensing surveys that mostly target K and M-dwarfs to the fraction of A and F-type stars with Neptune-mass planets beyond the snow-line to test our hypothesis that many hot white dwarfs are evaporating giant planets, and accrete some of their atmospheric material. 

For this test we use only the single stars of \citet{barstowetal14-1}, as binarity affects the statistics for two reasons. On one hand, stellar winds from relatively close main sequence companions can cause metal pollution \citep{pyrzasetal12-1} which is then impossible to 
distinguish from accretion of evaporated planetary atmospheres. On the other hand, the formation of giant planets can be impeded by the presence of close and/or massive stellar companions. \citet{holbergetal13-1} list stellar companions for 20 of the 89 systems of \citet{barstowetal14-1}, however, this compilation ignored M-type companions. Scrutiny of the literature reveals one additional wide binary (WD\,0501+524) and three short-period white dwarf plus M-dwarf binaries. We also identified five new common proper motion companions using a two arc-min search within the \textit{Gaia} Data Release~2 \citep{gaiaetal18-1}. Finally, we removed WD\,0802+413 and WD\,0621$-$376 from the sample. 
While there is no clear evidence that they are part of binaries, their masses are too low to have evolved as single stars within the age of the Galaxy.

Using the final sample of 58 white dwarfs, we then performed Monte-Carlo simulations to test the suggested scenario.
%to test our hypothesis  
%Using the 58 single stars from  \citet{barstowetal14-1} 
We constructed a planet population model for the white dwarf progenitor stars with initial semi-major axes of $3-30$\,au. For the probability distributions of planet mass ratio and separation we used the equations and best fit values from recent microlensing planet surveys \citep[][their Table\,4]{suzukietal16-1}. 

The planet positions after the evolution of the host star into a white dwarf were determined assuming adiabatic mass loss. We computed the EUV luminosity of each white dwarf as outlined in Sect.\,\ref{s-solsys}. We then calculated for each white dwarf the mass accretion rate of evaporated material and evaluated whether it exceeds the detection limit of $10^5\,\mathrm{g\,s^{-1}}$. We furthermore randomly selected 25 per cent of the white dwarfs as being in addition metal polluted by rocky planetary debris as indicated by the observations of cool white dwarfs \citep{koesteretal14-1}. The only parameter we adjusted was the occurrence rate of planets with hydrogen-rich atmospheres. We find that the fraction of white dwarfs accreting detectable amounts of volatiles from planetary atmospheres evolves with cooling age, and hence effective temperature, consistent with the observations if at least $\simeq50$~per cent of the hot white dwarfs in the sample host at least one giant planet. Figure\,\ref{f-sim_obs_comp} shows the observed fraction of polluted hot white dwarfs and the predicted fraction of white dwarfs accreting from evaporating giant planets with hydrogen-rich atmospheres as a function of white dwarf temperature for a planet occurrence rate of 65~per cent. The remarkable agreement shows that evaporation of giant planets, unavoidable in the future solar system, offers a consistent explanation for the volatiles frequently detected in hot white dwarfs. 

Determining the detailed abundances of the accreted material will require additional work in the treatment of radiative levitation and systematic deep surveys of a well defined sample of hot white dwarfs, but offers the potential to infer the composition of extra-solar giant planets. 

\acknowledgements{The authors are grateful to Detlev Koester for sharing his model atmosphere code. MRS and FL are members of the Millennium Nuclues for Planet Formation (NPF). MRS also thanks for support from FONDECYT (grant 1181404). OT was supported by a Leverhulme Trust Research Project Grant. BTG and OT were supported by the UK STFC grant ST/P000495.}

%\bibliographystyle{aasjournal}
%\bibliography{aamnem99,ref}

\begin{thebibliography}{}
\expandafter\ifx\csname natexlab\endcsname\relax\def\natexlab#1{#1}\fi
\providecommand{\url}[1]{\href{#1}{#1}}

\bibitem[{{Ayres}(1997)}]{ayres97-1}
{Ayres}, T.~R. 1997, Journal of Geophysical Research, 102, 1641

\bibitem[{{Barstow} {et~al.}(2014){Barstow}, {Barstow}, {Casewell}, {Holberg},
  \& {Hubeny}}]{barstowetal14-1}
{Barstow}, M.~A., {Barstow}, J.~K., {Casewell}, S.~L., {Holberg}, J.~B., \&
  {Hubeny}, I. 2014, MNRAS, 440, 1607

\bibitem[{{Barstow} {et~al.}(2003){Barstow}, {Good}, {Holberg}, {Hubeny},
  {Bannister}, {Bruhweiler}, {Burleigh}, \& {Napiwotzki}}]{barstowetal03-1}
{Barstow}, M.~A., {Good}, S.~A., {Holberg}, J.~B., {et~al.} 2003, MNRAS, 341,
  870

\bibitem[{{Ben-Jaffel} \& {Ballester}(2013)}]{ben-jaffel+ballester13-1}
{Ben-Jaffel}, L., \& {Ballester}, G.~E. 2013, A\&A, 553, A52

\bibitem[{{Borgniet} {et~al.}(2019){Borgniet}, {Lagrange}, {Meunier}, {Galland
  }, {Arnold}, {Astudillo-Defru}, {Beuzit}, {Boisse}, {Bonfils}, \&
  {Bouchy}}]{borgnietetal19-1}
{Borgniet}, S., {Lagrange}, A.~M., {Meunier}, N., {et~al.} 2019, A\&A, 621, A87

\bibitem[{{Bourrier} {et~al.}(2018){Bourrier}, {Lecavelier des Etangs},
  {Ehrenreich}, {Sanz-Forcada}, {Allart}, {Ballester}, {Buchhave}, {Cohen},
  {Deming}, {Evans}, {Garc{\'{\i}}a Mu{\~n}oz}, {Henry}, {Kataria}, {Lavvas},
  {Lewis}, {L{\'o}pez-Morales}, {Marley}, {Sing}, \&
  {Wakeford}}]{bourrieretal18-1}
{Bourrier}, V., {Lecavelier des Etangs}, A., {Ehrenreich}, D., {et~al.} 2018,
  A\&A, 620, A147

\bibitem[{{Brown} {et~al.}(2017){Brown}, {Veras}, \&
  {G{\"a}nsicke}}]{brownetal17-1}
{Brown}, J.~C., {Veras}, D., \& {G{\"a}nsicke}, B.~T. 2017, \mnras, 468, 1575

\bibitem[{{Bruhweiler} \& {Kondo}(1981)}]{bruhweiler+kondo81-1}
{Bruhweiler}, F.~C., \& {Kondo}, Y. 1981, ApJ Lett., 248, L123

\bibitem[{{Cassan} {et~al.}(2012){Cassan}, {Kubas}, {Beaulieu}, {Dominik},
  {Horne}, {Greenhill}, {Wambsganss}, {Menzies}, {Williams}, {J{\o}rgensen},
  {Udalski}, {Bennett}, {Albrow}, {Batista}, {Brillant}, {Caldwell}, {Cole},
  {Coutures}, {Cook}, {Dieters}, {Dominis Prester}, {Donatowicz}, {Fouqu{\'e}},
  {Hill}, {Kains}, {Kane}, {Marquette}, {Martin}, {Pollard}, {Sahu}, {Vinter},
  {Warren}, {Watson}, {Zub}, {Sumi}, {Szyma{\'n}ski}, {Kubiak}, {Poleski},
  {Soszynski}, {Ulaczyk}, {Pietrzy{\'n}ski}, \& {Wyrzykowski}}]{cassanetal12-1}
{Cassan}, A., {Kubas}, D., {Beaulieu}, J.~P., {et~al.} 2012, Nat, 481, 167

\bibitem[{{Chauvin} {et~al.}(2017){Chauvin}, {Desidera}, {Lagrange}, {Vigan},
  {Gratton}, {Langlois}, {Bonnefoy}, {Beuzit}, {Feldt}, \&
  {Mouillet}}]{chauvinetal17-1}
{Chauvin}, G., {Desidera}, S., {Lagrange}, A.~M., {et~al.} 2017, A\&A, 605, L9

\bibitem[{{Chayer} {et~al.}(1995){Chayer}, {Vennes}, {Pradhan}, {Thejll},
  {Beauchamp}, {Fontaine}, \& {Wesemael}}]{chayeretal95-2}
{Chayer}, P., {Vennes}, S., {Pradhan}, A.~K., {et~al.} 1995, ApJ, 454, 429

\bibitem[{{Dickinson} {et~al.}(2013){Dickinson}, {Barstow}, \&
  {Welsh}}]{dickinsonetal13-1}
{Dickinson}, N.~J., {Barstow}, M.~A., \& {Welsh}, B.~Y. 2013, MNRAS, 428, 1873

\bibitem[{{Duncan} \& {Lissauer}(1998)}]{duncan+lissauer98-1}
{Duncan}, M.~J., \& {Lissauer}, J.~J. 1998, Icarus, 134, 303

\bibitem[{{Ehrenreich} {et~al.}(2012){Ehrenreich}, {Bourrier}, {Bonfils},
  {Lecavelier des Etangs}, {H{\'e}brard}, {Sing}, {Wheatley}, {Vidal-Madjar},
  {Delfosse}, {Udry}, {Forveille}, \& {Moutou}}]{ehrenreichetal12-1}
{Ehrenreich}, D., {Bourrier}, V., {Bonfils}, X., {et~al.} 2012, A\&A, 547, A18

\bibitem[{{Ehrenreich} {et~al.}(2015){Ehrenreich}, {Bourrier}, {Wheatley},
  {Lecavelier des Etangs}, {H{\'e}brard}, {Udry}, {Bonfils}, {Delfosse},
  {D{\'e}sert}, {Sing}, \& {Vidal-Madjar}}]{ehrenreichetal15-1}
{Ehrenreich}, D., {Bourrier}, V., {Wheatley}, P.~J., {et~al.} 2015, Nat, 522,
  459

\bibitem[{{Farihi}(2016)}]{farihi16-1}
{Farihi}, J. 2016, New Astronomy Reviews, 71, 9

\bibitem[{{Fontaine} {et~al.}(2001){Fontaine}, {Brassard}, \&
  {Bergeron}}]{fontaineetal01-1}
{Fontaine}, G., {Brassard}, P., \& {Bergeron}, P. 2001, PASP, 113, 409

\bibitem[{{Gaia Collaboration} {et~al.}(2018){Gaia Collaboration}, {Brown},
  {Vallenari}, {Prusti}, {de Bruijne}, {Babusiaux}, {Bailer-Jones}, {Biermann},
  {Evans}, {Eyer}, {Jansen}, {Jordi}, {Klioner}, {Lammers}, {Lindegren},
  {Luri}, {Mignard}, {Panem}, {Pourbaix}, {Randich}, \& et~al.}]{gaiaetal18-1}
{Gaia Collaboration}, {Brown}, A.~G.~A., {Vallenari}, A., {et~al.} 2018, A\&A,
  616, A1

\bibitem[{{Ghezzi} {et~al.}(2018){Ghezzi}, {Montet}, \&
  {Johnson}}]{ghezzietal18-1}
{Ghezzi}, L., {Montet}, B.~T., \& {Johnson}, J.~A. 2018, ApJ, 860, 109

\bibitem[{{Holberg} \& {Bergeron}(2006)}]{holberg+bergeron06-1}
{Holberg}, J.~B., \& {Bergeron}, P. 2006, AJ, 132, 1221

\bibitem[{{Holberg} {et~al.}(2013){Holberg}, {Oswalt}, {Sion}, {Barstow}, \&
  {Burleigh}}]{holbergetal13-1}
{Holberg}, J.~B., {Oswalt}, T.~D., {Sion}, E.~M., {Barstow}, M.~A., \&
  {Burleigh}, M.~R. 2013, MNRAS, 435, 2077

\bibitem[{{Jura}(2003)}]{jura03-1}
{Jura}, M. 2003, ApJ Lett., 584, L91

\bibitem[{{Jura} \& {Young}(2014)}]{jura+young14-1}
{Jura}, M., \& {Young}, E.~D. 2014, Annual Review of Earth and Planetary
  Sciences, 42, 45

\bibitem[{{Koester}(2010)}]{koester10-1}
{Koester}, D. 2010, Mem. Soc. Astron. Ital., 81, 921

\bibitem[{{Koester} {et~al.}(2014){Koester}, {G{\"a}nsicke}, \&
  {Farihi}}]{koesteretal14-1}
{Koester}, D., {G{\"a}nsicke}, B.~T., \& {Farihi}, J. 2014, A\&A, 566, A34

\bibitem[{{Marois} {et~al.}(2008){Marois}, {Macintosh}, {Barman}, {Zuckerman},
  {Song}, {Patience}, {Lafreni{\`e}re}, \& {Doyon}}]{maroisetal08-1}
{Marois}, C., {Macintosh}, B., {Barman}, T., {et~al.} 2008, Science, 322, 1348

\bibitem[{{Mayor} \& {Queloz}(1995)}]{mayor+queloz95-1}
{Mayor}, M., \& {Queloz}, D. 1995, Nat, 378, 355

\bibitem[{{Murray-Clay} {et~al.}(2009){Murray-Clay}, {Chiang}, \&
  {Murray}}]{murray-clayetal09-1}
{Murray-Clay}, R.~A., {Chiang}, E.~I., \& {Murray}, N. 2009, ApJ, 693, 23

\bibitem[{{Owen}(2019)}]{owen19-1}
{Owen}, J.~E. 2019, Annual Review of Earth and Planetary Sciences, 47, 67

\bibitem[{{Owen} \& {Alvarez}(2016)}]{owen+alvarez16-1}
{Owen}, J.~E., \& {Alvarez}, M.~A. 2016, ApJ, 816, 34

\bibitem[{{Owen} \& {Wu}(2017)}]{owen+wu17-1}
{Owen}, J.~E., \& {Wu}, Y. 2017, ApJ, 847, 29

\bibitem[{{Parker}(1964)}]{parker64-1}
{Parker}, E.~N. 1964, ApJ, 139, 72

\bibitem[{{Paxton} {et~al.}(2011){Paxton}, {Bildsten}, {Dotter}, {Herwig},
  {Lesaffre}, \& {Timmes}}]{paxtonetal11-1}
{Paxton}, B., {Bildsten}, L., {Dotter}, A., {et~al.} 2011, ApJS, 192, 3

\bibitem[{{Pizzolato} {et~al.}(2000){Pizzolato}, {Maggio}, \&
  {Sciortino}}]{pizzolatoetal00-1}
{Pizzolato}, N., {Maggio}, A., \& {Sciortino}, S. 2000, A\&A, 361, 614

\bibitem[{{Pyrzas} {et~al.}(2012){Pyrzas}, {G{\"a}nsicke}, {Brady}, {Parsons},
  {Marsh}, {Koester}, {Breedt}, {Copperwheat}, {Nebot G{\'o}mez-Mor{\'a}n},
  {Rebassa-Mansergas}, {Schreiber}, \& {Zorotovic}}]{pyrzasetal12-1}
{Pyrzas}, S., {G{\"a}nsicke}, B.~T., {Brady}, S., {et~al.} 2012, MNRAS, 419,
  817

\bibitem[{{Sanz-Forcada} {et~al.}(2011){Sanz-Forcada}, {Micela}, {Ribas},
  {Pollock}, {Eiroa}, {Velasco}, {Solano}, \&
  {Garc{\'{\i}}a-{\'A}lvarez}}]{sanz-forcadaetal11-1}
{Sanz-Forcada}, J., {Micela}, G., {Ribas}, I., {et~al.} 2011, A\&A, 532, A6

\bibitem[{{Schatzman}(1948)}]{schatzman48-1}
{Schatzman}, E. 1948, Nat, 161, 61

\bibitem[{{Shapiro} \& {Lightman}(1976)}]{shapiro+lightman76-1}
{Shapiro}, S.~L., \& {Lightman}, A.~P. 1976, ApJ, 204, 555

\bibitem[{{Shvartzvald} {et~al.}(2016){Shvartzvald}, {Maoz}, {Udalski}, {Sumi},
  {Friedmann}, {Kaspi}, {Poleski}, {Szyma{\'n}ski}, {Skowron}, {Koz{\l}owski},
  {Wyrzykowski}, {Mr{\'o}z}, {Pietrukowicz}, {Pietrzy{\'n}ski},
  {Soszy{\'n}ski}, {Ulaczyk}, {Abe}, {Barry}, {Bennett}, {Bhattacharya},
  {Bond}, {Freeman}, {Inayama}, {Itow}, {Koshimoto}, {Ling}, {Masuda}, {Fukui},
  {Matsubara}, {Muraki}, {Ohnishi}, {Rattenbury}, {Saito}, {Sullivan},
  {Suzuki}, {Tristram}, {Wakiyama}, \& {Yonehara}}]{shvartzvaldetal16-1}
{Shvartzvald}, Y., {Maoz}, D., {Udalski}, A., {et~al.} 2016, MNRAS, 457, 4089

\bibitem[{{Suzuki} {et~al.}(2016){Suzuki}, {Bennett}, {Sumi}, {Bond}, {Rogers},
  {Abe}, {Asakura}, {Bhattacharya}, {Donachie}, {Freeman}, {Fukui}, {Hirao},
  {Itow}, {Koshimoto}, {Li}, {Ling}, {Masuda}, {Matsubara}, {Muraki},
  {Nagakane}, {Onishi}, {Oyokawa}, {Rattenbury}, {Saito}, {Sharan}, {Shibai},
  {Sullivan}, {Tristram}, {Yonehara}, \& {MOA Collaboration}}]{suzukietal16-1}
{Suzuki}, D., {Bennett}, D.~P., {Sumi}, T., {et~al.} 2016, ApJ, 833, 145

\bibitem[{{Tian} {et~al.}(2005){Tian}, {Toon}, {Pavlov}, \& {De
  Sterck}}]{tianetal05-1}
{Tian}, F., {Toon}, O.~B., {Pavlov}, A.~A., \& {De Sterck}, H. 2005, ApJ, 621,
  1049

\bibitem[{{Tripathi} {et~al.}(2015){Tripathi}, {Kratter}, {Murray-Clay}, \&
  {Krumholz}}]{tripathietal15-1}
{Tripathi}, A., {Kratter}, K.~M., {Murray-Clay}, R.~A., \& {Krumholz}, M.~R.
  2015, ApJ, 808, 173

\bibitem[{{Tu} {et~al.}(2015){Tu}, {Johnstone}, {G{\"u}del}, \&
  {Lammer}}]{tuetal15-1}
{Tu}, L., {Johnstone}, C.~P., {G{\"u}del}, M., \& {Lammer}, H. 2015, A\&A, 577,
  L3

\bibitem[{{Veras}(2016)}]{veras16-1}
{Veras}, D. 2016, MNRAS, 463, 2958

\bibitem[{{Vidal-Madjar} {et~al.}(2003){Vidal-Madjar}, {Lecavelier des Etangs},
  {D{\'e}sert}, {Ballester}, {Ferlet}, {H{\'e}brard}, \&
  {Mayor}}]{vidal-madjaretal03-1}
{Vidal-Madjar}, A., {Lecavelier des Etangs}, A., {D{\'e}sert}, J.-M., {et~al.}
  2003, Nat, 422, 143

\bibitem[{{Vidal-Madjar} {et~al.}(2004){Vidal-Madjar}, {D{\'e}sert},
  {Lecavelier des Etangs}, {H{\'e}brard}, {Ballester}, {Ehrenreich}, {Ferlet},
  {McConnell}, {Mayor}, \& {Parkinson}}]{vidal-madjaretal04-1}
{Vidal-Madjar}, A., {D{\'e}sert}, J.-M., {Lecavelier des Etangs}, A., {et~al.}
  2004, ApJ Lett., 604, L69

\bibitem[{{Wang}(1981)}]{wang81-1}
{Wang}, Y.-M. 1981, A\&A, 102, 36

\bibitem[{{Xu} {et~al.}(2014){Xu}, {Jura}, {Koester}, {Klein}, \&
  {Zuckerman}}]{xuetal14-1}
{Xu}, S., {Jura}, M., {Koester}, D., {Klein}, B., \& {Zuckerman}, B. 2014, ApJ,
  783, 79

\bibitem[{{Yelle} {et~al.}(2008){Yelle}, {Lammer}, \& {Ip}}]{yelleetal08-1}
{Yelle}, R., {Lammer}, H., \& {Ip}, W.-H. 2008, Space Science Reviews, 139, 437

\bibitem[{{Zuckerman} {et~al.}(2003){Zuckerman}, {Koester}, {Reid}, \&
  {H{\"u}nsch}}]{zuckermanetal03-1}
{Zuckerman}, B., {Koester}, D., {Reid}, I.~N., \& {H{\"u}nsch}, M. 2003, ApJ,
  596, 477

\end{thebibliography}

\end{document}